\documentclass[amsmath,amssymb,14pt]{article}

\usepackage{graphicx, epsfig}

\usepackage{amssymb,amsmath,color}

\usepackage{subcaption}
\usepackage{caption}
\usepackage[export]{adjustbox}

\usepackage[nottoc]{tocbibind}

\usepackage{dcolumn}
\usepackage{bm}
\newtheorem{mytheo*}{Theorem}[section]
\newtheorem{mylemma*}{Lemma}[section]
\newtheorem{myproposition*}{Proposition}[section]
\newtheorem{mydef*}{Definition}[section]
\newtheorem{myremark*}{Remark}[section]

\usepackage{mathtools}

\usepackage{authblk}

\begin{document}
\title{\textsc{A Direct Method For the Low Energy Scattering Solution of Delta Shell Potentials}}

\author[1]{Fatih Erman}
\author[2]{Sema Seymen}
\affil[1]{Department of Mathematics, \.{I}zmir Institute of Technology, Urla, 35430, \.{I}zmir, Turkey}
\affil[2]{Department of Physics, Bo\u{g}azi\c{c}i University, Bebek, 34342, \.{I}stanbul, Turkey}
\affil[ ]{e-mail: fatih.erman@gmail.com}

\maketitle

\begin{abstract}
A direct method for the bound states and the low energy scattering from a circular and a spherical delta shell potentials is proposed and the results are compared with the one using the standard partial wave analysis developed for potentials with rotational symmetry. The formulation is presented in momentum space and the scattering solutions are obtained by considering the elementary use of distributions. In this approach, the outgoing boundary conditions are imposed explicitly in contrast to the  $i\epsilon$ prescription often used in quantum mechanics.   
\end{abstract}


\section{Introduction}

One of the main goals in elementary quantum mechanics is to find  solutions of the time-independent Schr\"{o}dinger equation in $\mathbb{R}^3$
\begin{eqnarray} \label{scheq}
 -\nabla^2 \psi(\mathbf{r}) + V(\mathbf{r}) \psi(\mathbf{r}) = E \psi(\mathbf{r}) \;,
\end{eqnarray}
for a given potential $V$, where units are chosen such that $\hbar=2m=1$ for simplicity. The left hand side of this equation is simply written as the action of a self-adjoint operator $H$, called Hamiltonian, on wave functions $\psi$. The decaying square integrable solutions to Equation (\ref{scheq}) are known as the bound state wave functions and the values of $E$ satisfying (\ref{scheq}) are called the bound state energies. These solutions physically represent the cases where a particle is localized in a region of space by some potential well $V$.

Another class of solutions correspond to a scattering experiment, where a beam of particles with a definite momentum $\mathbf{p}=\mathbf{k}$ (and hence a definite energy $E=k^2$)   hits a target particle and the distribution of impinging particles going away from  the target one is studied. Although the actual physical problem is time-dependent, as a typical situation involves a wave packet moving in and then scattering  off to infinity,  it is usually sufficient to study the stationary scattering problem of a particle with a definite momentum. Moreover, in realistic situations, there are multiple scattering events that one has to take into account as well as inelastic events, those that   actually exchange energy with  the target particle. Nevertheless, the essence of the problem is captured by the present simpler version, elastic scattering from a potential.

For a wave packet, once we find the 
particular solution to the stationary problem satisfying a specific asymptotic boundary condition (known as the outgoing boundary condition, or also called Sommerfeld radiation condition) \cite{griffiths2016introduction, faddeev2009lectures}
\begin{eqnarray}
\psi(\mathbf{r}, \mathbf{k}) = 
N\left( e^{i \mathbf{k} \cdot \mathbf{r}} + f(k, \theta) \frac{e^{i k r}}{r} + o\left(\frac{1}{r}\right)\right) \;,
\label{outgoingboundaryconditions}
\end{eqnarray}
for large $r>0$, and determine the scattering amplitude $f$,  the time-dependent wave packet solution to the time-dependent Schr\"{o}dinger equation ($i \frac{\partial \Psi}{\partial t}= H \Psi$) can be obtained as 
\begin{eqnarray}
\Psi(\mathbf{r},t)= \int_{\mathbb{R}^3} C(\mathbf{k}) \psi(\mathbf{r}, \mathbf{k}) e^{-i k^2 t} \frac{d^3 k}{(2\pi)^3} \;.
\end{eqnarray}
The scattering information does not depend on the form of the function $C(\mathbf{k})$, moreover,  as long as this function  is concentrated around a  particular $\mathbf{k}$ the above picture becomes more accurate (see Ref. \cite{faddeev2009lectures}, pages 175-180). 

Note that the scattering wave function $\psi$ depends on both $\mathbf{r}$ and $\mathbf{k}$, from now on we shall prefer to write $\psi(\mathbf{r})$ for simplicity. 
The first term in the above boundary condition corresponds asymptotically to the incoming particles with  a definite direction $\mathbf{k}$ and the second term  corresponds to the scattered particle moving away from the center and the form of its wave function depends on the angle $\theta$ only (if the potential is assumed to be spherically symmetric, i.e., Hamiltonian commutes with the angular momentum operators). In this case,  it is convenient to use the so-called partial wave analysis \cite{griffiths2016introduction, shankar2012principles}, where the Schr\"{o}dinger equation admits a separable solution whose radial part can be reduced to an effectively one-dimensional problem containing a term with angular momentum quantum number $l$.

Since we will study a particular class of a singular potentials in one, two and three dimensions, let us briefly comment on the one-dimensional and two-dimensional scattering problem as well. For the one-dimensional case, there are two asymptotic regions, namely the far left (say region I) and far right (say region II). The outgoing boundary condition then states that the wave function must go like $e^{i k x}$ in the region $II$ and $e^{-i k x}$ in the region $I$. The wave function for the incoming particle is either $e^{ik x}$ or $e^{-ik x}$, depending on which direction the particle is sent to the target (see Ref. \cite{shankar2012principles}, pages 176-173 and Ref. \cite{faddeev2009lectures}, pages 159-167). In this case, the scattering problem is much simpler as expected.
For the two-dimensional case, the outgoing boundary condition given in (\ref{outgoingboundaryconditions}) must be replaced by \cite{lapidus1982quantum}
\begin{eqnarray}
\psi(\mathbf{r}, \mathbf{k}) = 
N\left( e^{i \mathbf{k} \cdot \mathbf{r}} + f(k, \theta) \frac{e^{i k r}}{\sqrt{r}} + o\left(\frac{1}{\sqrt{r}}\right)\right) \;.
\label{outgoingboundaryconditions2d}
\end{eqnarray}

To clarify our approach, it is a good exercise to go over  the Delta-function potential in one dimension
\begin{eqnarray}
V(x)=-\lambda \delta(x) \;, \label{deltapotential1d}
\end{eqnarray}
where the constant $\lambda>0$ is called the strength of the interaction (and assumed to be positive for an attractive case). This potential ideally represents the situation where the range of a potential is negligible compared to the de Broglie wavelength of the particle under consideration (see, Ref. \cite{griffiths2016introduction}, pages 70-75 for its standard bound state and scattering solutions in position space). 
It is important to emphasize that this way of writing the Hamiltonian for delta potentials is not mathematically rigorous since the above formal expression of the operator does not send square integrable functions into square integrable functions. One way to deal with this issue is based on the idea of Von Neumann's self-adjoint extension theory (see Refs. \cite{bonneau2001self, araujo2004operator} for the pedagogical introduction to the subject and Refs. \cite{albeverio2000singular, albeverio2012solvable, exner2015quantum} for the detailed expositions of such singular potentials). Nevertheless, the formal operator approach without taking into account  these  delicate mathematical issues will not affect the spectrum  in the end. Therefore we   study the bound state and scattering problems for such singular potentials in a formal manner. For this reason, we first write the  Hamiltonian operator associated with the above potential (\ref{deltapotential1d}) as    
\begin{eqnarray}
H= H_0 - \lambda | 0 \rangle \langle 0 | \;, \label{formaldeltapotential}
\end{eqnarray}
where the interaction term is expressed as a projection operator onto the generalized position ket $|x=0 \rangle$ (see Ref. \cite{appel2007mathematics, bohm2013quantum} for the definition of generalized kets and bras).  This equivalence can be seen formally by noticing the property of Dirac delta functions $\langle x | 0 \rangle \langle 0|\psi \rangle= \delta(x) \psi(0)=\delta(x) \psi(x)$. One may think of the above formal expression as some kind of limit of the regularized versions of the Hamiltonian. We shall use the word ``formal" throughout the paper to indicate that we only consider the form of the equations without  worrying much about their rigorous mathematical aspects.

Finding the solution of bound states or the scattering states in momentum space
 is in general not easier than finding the solution of the original problem (\ref{scheq}) in position space. However, for  Dirac delta potentials in one dimension we have, 
 \begin{eqnarray}
 \langle p |H | \psi \rangle = p^2 \widehat{\psi}(p) - \lambda \psi(0) = E \widehat{\psi}(p) \;, \label{algebraicequation}
 \end{eqnarray}
and the problem is reduced to finding the solution of an algebraic equation \cite{lieber1975quantum}. Here $\widehat{\psi}(p)$ is the Fourier transform of $\psi(x)$ and $\psi(0)$ corresponds to  the value of the wave function at the location of the delta function\footnote{Incidentally, for a bound state solution, which has well defined kinetic energy, one can pick a continuous function to represent the wave function and it makes sense to talk about its value at the origin.}. The stationary scattering solution to the above algebraic equation for $E=k^2$ can be found by using the so-called $i \epsilon$ prescription \cite{shankar2012principles, erman2018scattering} or alternatively by taking into account the distributional solution to these algebraic equation (\ref{algebraicequation}), as discussed  in \cite{lieber1975quantum}. A similar distributional approach has been used in finding the fundamental solution of Helmholtz equation \cite{schmalz2010derivation}, as well. For the convenience of the reader, we will first review the Dirac delta potential in one dimension within this distributional approach and then extend it to higher dimensions. This alternative approach will be our main focus in this work. As we will see, the homogenous distributional solutions for two/three dimensional radially symmetric Dirac delta potential case will include extra terms essentially due to the fact that the Dirac delta functions in higher dimensions satisfy both $(p^2-k^2)\delta(\mathbf{p}\pm \mathbf{k})=0$ and $(p^2-k^2)\delta(p-k)=0$ in contrast to the one-dimensional case, where we have only $(p^2-k^2)\delta(p\pm k)=0$.

As possible generalizations, one can consider finitely many \cite{lapidus1982quantum, erman2018scattering} or periodically located  Dirac delta potentials, one of the simplest models in solid state physics for describing the band structure of metals, known as the Kronig-Penney model (see Ref. \cite{ashcroft1976solid}, page 149). Further applications in several areas of physics have been discussed in the book \cite{demkov2013zero} and some other interesting aspects of these potentials  have been recently presented in a review \cite{belloni2014infinite}.

In this work, we shall concentrate on the low energy solutions of the radially symmetric Dirac delta potentials in two and three dimensions. The spherical delta shell potential in three dimensions is one of the most well known potential in the literature, and given explicitly by  
\begin{eqnarray}
V=-\lambda \delta(r-R) \;. \label{deltashellpotential3d}
\end{eqnarray}
Since this is a spherically symmetric potential, it is amenable to make partial wave analysis commonly used in scattering problems (see, e.g. \cite{griffiths2016introduction}). As mentioned briefly above, the idea  is based on the fact that the rotational symmetry (Hamiltonian commutes with the angular momentum operator) allows us to split the problem into simpler pieces labeled by the angular momentum quantum number $l$. The scattering solutions of this  problem via  the partial wave analysis is summarized in Appendix A in order to make the paper as  self-contained as possible. There is a similar  analysis for the two dimensional version of this problem, delta function supported on a circle, which we  present in another section below, and its partial wave analysis is summarized in Appendix B.

This work is  an attempt to solve directly in momentum space the bound state and scattering states  of delta shell type of potentials without going into partial wave analysis. The direct formal  approach, in contrast to the differential equation approach,  provides a clear picture of the solutions. For this reason, in analogy with the point like Dirac delta potential expressed by (\ref{formaldeltapotential}), we propose the following formal Hamiltonian operator
\begin{eqnarray} \label{hamiltoniansphere}
    H= H_0 - \lambda  |\delta_{S^2} \rangle \langle \delta_{S^2} |  \;,
\end{eqnarray}
as a candidate for the spherical delta shell potential, where $H_0$ is the free Hamiltonian and the potential term corresponds to the interaction of a single particle with an external shell-like Dirac delta  potential. 
The Dirac delta function $\delta_{S}$ supported on the sphere $S^2$ is most naturally  defined by its action on test functions $\psi$,  (see Ref. \cite{appel2007mathematics}, page 195)
\begin{eqnarray}
\langle \delta_{S^2}| \psi \rangle & = & \frac{1}{\sqrt{A(S^2)}} \int_{S^2} \psi \; d S \;, 
\end{eqnarray}
where  $d S =R^2 d \Omega= R^2 \sin \theta d \theta d \phi$ and $A(S^2)=4 \pi R^2$. Here the bracket $\langle \cdot | \cdot \rangle$ should  be understood in the sense of the action of the Dirac delta distribution on the test functions\footnote{This is a delicate issue, we write the integral of the wave function that we are after, it is not obvious that the resulting wave function  is actually a  test function. Of course, in this particular example we only need the restriction of the wave function onto the sphere to be integrable, this turns out to be true, due to some nontrivial results in distributions, moreover the sphere being a compact space is of importance.}. Division by the square root of the surface area of the sphere is due to a normalization convention we introduce, which can be interpreted as the strength being proportional to the average of the wave function on the sphere. We note that this definition of the  Hamiltonian  corresponds directly to the $l=0$ sector of the standard delta shell potential given in (\ref{deltashellpotential3d}) in the partial wave analysis. Here, we  study the distributional solution of the scattering problem for the Hamiltonian (\ref{hamiltoniansphere}) in momentum space by extending the  ideas developed for the point Dirac delta potentials  in \cite{lieber1975quantum}. The advantage of this method is that the boundary conditions are explicitly used in finding the scattering solutions in contrast to the $i \epsilon$ prescription, where the boundary conditions are implicitly used. We also study the circular delta shell potential within the same approach for the sake of completeness--as  two dimensional materials gaining more importance in applications, we believe this will be an instructive example.

It should be  possible to extend the present approach to the point like Dirac delta potentials in higher dimensions. However, this problem requires renormalization (see Refs. \cite{mead1991analytical, jackiw1995diverse, mitra1998regularization, gosdzinsky1991learning, manuel1994perturbative, nyeo2000regularization}) so it exceeds the scope of the present work. Higher dimensional extension of spherical shell potentials have been also discussed in \cite{demiralp2003properties} and various extension of spherical shell delta potentials in a more mathematically context has been recently discussed in \cite{Fassari}, but we do  not develop them here either. More recently, an exact solution for circular delta potential in position space based on the calculation of eigenfunctions and eigenvalues of a certain relevant integral operator is given and results are compared with numerical results in \cite{maioli2018exact}. A similar work for the solution of the scattering of a plane wave from a spherical shell potential with angular dependent coupling strengths is given in \cite{azado2021quantum}.

The paper is organized as follows: In Section \ref{Point Dirac Delta Potential in One Dimensional Momentum Space}, we briefly sketch the distributional solutions of of point Dirac delta potentials in one dimension, which was originally discussed in \cite{lieber1975quantum}. In Section \ref{Spherical Dirac Delta Shell Potential in Momentum Space}, we introduce the delta potentials supported by sphere in three dimensions as a projection operator and then proceed with the study of its bound states and scattering states within the same distributional approach. Two dimensional  version of this problem is discussed in Section \ref{Circular Dirac Delta Potential in Momentum Space}. Finally, we give a short explanation why the formal operator approach corresponds to the  result of $l=0$ sector when  the delta shell potential is treated via differential equations. Appendices A and B are devoted to review the delta potentials supported by circle and sphere within the partial wave analysis. Appendix C contains a short proof for the distributional solutions of algebraic equations within the spirit of the paper.

\section{Point Dirac Delta Potential in One Dimensional Momentum Space}
\label{Point Dirac Delta Potential in One Dimensional Momentum Space}

The Dirac delta potential in one dimension is the most well-known textbook example for exactly solvable potentials \cite{griffiths2016introduction}. The standard way of solving the bound state and scattering problem is to separate the real line into two regions determined by the support of the delta potential, namely at $x=0$ and find the general solution to the time-independent Schr\"{o}dinger equation in each region. For the bound state problem, we should eliminate the solutions that blow up  as $x \rightarrow \pm \infty$. Then, imposing the continuity of the wave function at $x=0$ and relating the jump in the derivative of the wave function there to the wave function itself, we find the bound state energy and the associated wave function (see Ref. \cite{griffiths2016introduction}). As for the scattering problem, the general solution in each   region is glued by the continuity and the jump  of the derivative of wave function at $x=0$ in a similar way, then  we impose a natural boundary condition for the scattering phenomena (no reflection term in the positive $x$ axis if we assume that the particle is sent from the leftmost region) to get the reflection and transmission coefficients \cite{griffiths2016introduction}.

This same problem can formally be solved in momentum space as well, as discussed in \cite{lieber1975quantum}. Let us first briefly review this momentum space approach to the bound state and distributional solution of the scattering problem in one dimension below.

\subsection{Bound State problem}

We start with the time-independent Sch\"{o}dinger equation with the attractive formal Dirac delta potential $V(x)=-\lambda \delta(x)$ and parametrize the energy by $E=-\nu^2$ for simplicity. As mentioned in the  Introduction, we find in momentum space
\begin{eqnarray}
(p^2 + \nu^2) \widehat{\psi}(p)= \lambda \psi(0) \;. \label{boundstatewavefuncmomentum1d}
\end{eqnarray}
Dividing both sides of this algebraic equation by $p^2 + \nu^2$, we find the bound state wave function in momentum space 
\begin{eqnarray}
\widehat{\psi}(p)= \frac{\lambda \psi(0)}{p^2 +\nu^2} \;, \label{momentumspaceboundstatesolution1d}
\end{eqnarray}
which includes the unknown complex number $\psi(0)$ and $\nu$. The factor $\psi(0)$ is not essential since it can be absorbed into the normalization constant\footnote{As remarked previously, this makes sense  in one dimension. Moreover a bound state must be square integrable, as this form of the wave function in Fourier space confirms.}. To find $\nu$, let us impose the following consistency condition (the inverse Fourier transform of the above wave function (\ref{momentumspaceboundstatesolution1d}) evaluated at $x=0$ must be $\psi(0)$), that is,  
\begin{eqnarray}
\psi(0)= \int_{-\infty}^{\infty} \widehat{\psi}(p) \frac{dp}{2\pi} =   \int_{-\infty}^{\infty} \frac{\lambda \psi(0)}{p^2 + \nu^2} \frac{dp}{2\pi} \;. \label{consistency1d}
\end{eqnarray}
Since $\psi(0) \neq 0$ (otherwise the identity $\delta(x)\psi(x)=\delta(x)\psi(0)$ implies that the delta interaction is absent in the Hamiltonian), we find the bound state energy by evaluating the above elementary integral and solving $\nu$ 
\begin{eqnarray}
E=-\frac{\lambda^2}{4} \;.
\end{eqnarray}
It follows from the consistency condition (\ref{consistency1d}) that the bound state energy exists as long as $\lambda>0$. The bound state wave function can then be easily found by taking the inverse Fourier transformation of $\widehat{\psi}(p)$ given in Equation (\ref{momentumspaceboundstatesolution1d}). One  can compute this inverse transform  (for example by residue theorem \cite{brown2009complex}) to get the normalized bound state wave function
\begin{eqnarray}
\psi(x)= \sqrt{\frac{\lambda}{2}} \; e^{-\frac{\lambda}{2}|x|} \;.
\end{eqnarray}
This result is of course well-known in the literature \cite{griffiths2016introduction}.

\subsection{Stationary Scattering Problem}

Fourier transform of the time-independent Schr\"{o}dinger equation for $E=k^2$ yields
\begin{eqnarray}
(p^2-k^2) \widehat{\psi}(p)= \lambda \psi(0) \;. \label{algebraiceqscattering1d}
\end{eqnarray}
In contrast to the bound state case, we can not divide here both sides by $p^2-k^2$, which vanishes  at $p=\pm k$ since this leads to  singularities. The solution obtained in momentum space must be converted back to the position space by an inverse Fourier transform and these  singularities may cause trouble. The standard way of handling this problem is based on the idea of regularization by adding first a small complex term $-i \epsilon$ to $(p^2-k^2)$, thus removing the poles from the path of integration and then consider the limit $\epsilon \rightarrow 0^+$ at the end. Adding this complex term $-i \epsilon$ corresponds to the outgoing boundary condition \cite{shankar2012principles}. Another resolution of this problem is based on the idea that we look for solutions by means of the so-called generalized functions or distributions \cite{appel2007mathematics}. Distributions are continuous linear functionals acting on functions (more precisely linear mappings from the set of sufficiently smooth functions, called test functions, into the real numbers). The condition of continuity of these linear functionals  is a somewhat too technical issue for the present work,  we only refer the reader to Ref. \cite{appel2007mathematics} for some details  and for a more complete rigorous formulation of continuity  see also Ref. \cite{kanwal1998generalized}. There are two important classes of distributions. One class is known as regular distributions denoted by $T$, whose action on the test functions $\psi$ can be written as $\langle T, \psi \rangle = \int_{\mathbb{R}} f(x) \psi(x) d x$ for some locally integrable function $f(x)$ (see Ref. \cite{appel2007mathematics}, page 184 for the details). The principal value of $1/x$ is an example of a regular distribution, defined by
\begin{eqnarray}
\langle \mathrm{pv} \frac{1}{x}, \psi \rangle = \mathrm{pv} \int_{\mathbb{R}} \frac{\psi(x)}{x} dx = \lim_{\epsilon \rightarrow 0^+} \int_{|x|>\epsilon} \frac{\psi(x)}{x} d x \;.
\end{eqnarray}
It is important to notice that there is a ambiguity in the notation, where we use the same notation for the principal value distribution $1/x$ and the Cauchy principal value of the integral. All the other distributions which can not be written as above are called singular distributions. Simple example is the well-known point Dirac delta distribution, defined by $\langle \delta, \psi \rangle = \psi(0)$. It is customary to write formally $\langle \delta, \psi \rangle= \int_{\mathbb{R}} \delta(x) \psi(x) dx = \psi(0)$ in textbooks in quantum mechanics \cite{griffiths2016introduction, shankar2012principles}. One must always think of the distributions as some objects which takes test functions as their input and give real numbers as their output. For the sake of simplicity, we follow the same notation used in the  physics literature. However, whenever we write the expressions involving distributions in this paper, one must think of them as if they act on some test functions. Fourier transform of distributions $T$ acting on smooth functions which decay faster than any inverse power of $x$ are defined by $\langle \mathcal{F}(T), \psi \rangle = \langle T, \mathcal{F}(\psi) \rangle$ (see \cite{appel2007mathematics}). The most important result from our point of view in distribution theory is that if we have an algebraic equation 
\begin{eqnarray}
(x^2-a^2)T(x)=0 \;, 
\end{eqnarray}
where $a>0$, the general distributional solution (which includes singular ones) is formally given \cite{appel2007mathematics} by 
\begin{eqnarray}
T(x)= A \delta(x-a) + B \delta(x+ a) + \mathrm{pv} \left(\frac{1}{x^2-a^2}\right) \;, \label{distsolution1d}
\end{eqnarray}
where $A$ and $B$ are arbitrary complex numbers and  
\begin{eqnarray}
\mathrm{pv} \left( \frac{1}{x^2-a^2} \right) =   \frac{1}{2a} \Bigg( \mathrm{pv} \left( \frac{1}{x-a}\right) -\mathrm{pv} \left( \frac{1}{x+a}\right) \Bigg) \;. \label{pvp2k2}
\end{eqnarray}
The proof of this result is given in \cite{appel2007mathematics} and we briefly sketch it in Appendix C.

Therefore, using this result (\ref{distsolution1d}), the general solution of Equation (\ref{algebraiceqscattering1d}) is given by
\begin{eqnarray}
\widehat{\psi}(p)= A \delta(p-k)+ B \delta(p+k) + \lambda \psi(0) \mathrm{pv} \left( \frac{1}{p^2-k^2} \right) \;,
\end{eqnarray}
where the delta functions appear in the first two terms since they correspond to the homogeneous part of equation $(p^2-k^2)\widehat{\psi}(p)=0$. 
Substituting  this back into inverse Fourier transformation, we find 
\begin{eqnarray}
\psi(x)= \frac{A}{2\pi} e^{i k x} + \frac{B}{2\pi} e^{-i k x} + \lambda \psi(0) \; \mathrm{pv} \int_{-\infty}^{\infty} \frac{e^{i p x}}{p^2-k^2} \; \frac{d p}{2\pi} \;.
\end{eqnarray}
Let us first find the following principal value 
\begin{eqnarray}
I(x,k)= \mathrm{pv} \int_{-\infty}^{\infty} \frac{e^{i p x}}{p-k} \; \frac{d p}{2\pi} \;.
\end{eqnarray}
This can be easily evaluated by using the residue theorem. For $x>0$, one can choose the contour consisting of the real axis going around the pole $p=k$ symmetrically along a semicircle of radius $\epsilon$ and the semicircle of radius $R$ in the upper half plane. Then, in the limit as $\epsilon \rightarrow 0^+$ and $R \rightarrow \infty$ together with the Jordan's lemma, we end up with $I(x,k)= \frac{i}{2} e^{i k x}$ for $x>0$. Similarly, for $x<0$ we choose the contour consisting of the real axis going around the pole $p=k$ symmetrically along a semicircle of radius $\epsilon$ and the semicircle of radius $R$ in the lower half plane in this case. Then, we get $I(x,k)=-\frac{i}{2} e^{ik x}$. Hence, we find
\begin{eqnarray}
I(x,k)=  \begin{cases}
 - \frac{i}{2}  \; e^{i k x}  & x < 0 \;, \\  \frac{i}{2}  \; e^{i k x}  & x > 0 \;.
 \end{cases}
\end{eqnarray}
After substitution of this result in the solution $\psi(x)$ and using (\ref{pvp2k2}), we obtain 
\begin{eqnarray}
\psi(x)=  \begin{cases}
\frac{A}{2\pi} e^{i k x} + \frac{B}{2\pi} e^{-i k x} - \frac{i \lambda \psi(0)}{4k} \; \left( e^{i k x} - e^{-i k x}\right) & x<0 \;, \\ \frac{A}{2\pi} e^{i k x} + \frac{B}{2\pi} e^{-i k x} + \frac{i \lambda \psi(0)}{4k} \; \left( e^{i k x} - e^{-i k x}\right) & x>0 \;.
\end{cases}
\end{eqnarray}
By the continuity of the wave function at $x=0$, $\psi(0)= (A+B)/(2\pi)$. This leads to
\begin{eqnarray}
\psi(x)=
\begin{cases}
\frac{A}{2\pi} e^{i k x} + \frac{B}{2\pi} e^{-i k x} -\frac{i \lambda (A+B)}{8 \pi k} \left( e^{ik x}-e^{-ik x}\right) & \mathrm{for} \; x \leq 0 \;, \\ \\ 
\frac{A}{2\pi} e^{i k x} + \frac{B}{2\pi} e^{-i k x} + \frac{i \lambda (A+B)}{8 \pi k} \left( e^{ik x}-e^{-ik x}\right)  & \mathrm{for} \; x \geq 0 \;.
\end{cases}
\end{eqnarray}
Suppose the incoming particle is sent from the far negative $x$ axis. In this situation, we physically expect that there will be no reflection terms $e^{-ikx}$ in the far positive $x$ axis. This condition forces us to conclude that
\begin{eqnarray}
B= \frac{i \lambda}{4k-i \lambda} \; A \;.
\end{eqnarray}
Substituting this into the above solution we finally obtain
\begin{eqnarray}
\psi(x)=
\begin{cases}
\frac{A}{2\pi} \left( \frac{4k-2 i \lambda}{4k-i \lambda} \right) e^{i k x} + \frac{A}{2\pi} \left( \frac{2 i \lambda}{4k-i \lambda}\right) e^{-i k x} & \mathrm{for} \; x \leq 0 \;, \\ \\
 \frac{A}{2\pi} \left(1+ \frac{i \lambda}{4k-i \lambda} \right) e^{ik x}  & \mathrm{for} \; x \geq 0 \;.
\end{cases}
\end{eqnarray}
From this scattering solution, one can easily read the reflection and transmission coefficients
$T  =  \frac{4k^2}{4k^2 + \lambda^2}$ and $R = \frac{\lambda^2}{4k^2 + \lambda^2}$, which are the well-known results given in the literature \cite{griffiths2016introduction}.

\section{Spherical Dirac Delta Shell Potential in Three Dimensional Momentum Space}
\label{Spherical Dirac Delta Shell Potential in Momentum Space}

\subsection{Bound State Problem}

For the bound state problem of the spherical delta shell potential, we need to find the decaying square integrable solutions of the time-independent Schr\"{o}dinger equation associated with the Hamiltonian (\ref{hamiltoniansphere}). 
We can now follow analogously the one dimensional problem so that the Fourier transformation of the time-independent Schr\"{o}dinger equation for this potential is given by
\begin{eqnarray}
\left(p^2 + \nu^2 \right) \widehat{\psi}(\mathbf{p}) & = & \lambda \langle \mathbf{p} | \delta_{S^2} \rangle \langle \delta_{S^2}| \psi \rangle \nonumber \\
& = & \frac{\lambda}{4 \pi R^2} \left( \int_{S^2} e^{-i \mathbf{p}\cdot \boldsymbol{\sigma}} R^2 d \Omega \right) \left(\int_{S^2} \psi(\boldsymbol{\sigma}) R^2 d \Omega \right) \;, \label{scheqboundstateinmomentum}
\end{eqnarray}
where the sphere is covered by a single chart, and given by its well-known local parametrization $\boldsymbol{\sigma}:(0,2\pi) \times (0, \pi) \rightarrow S^2$:
\begin{eqnarray}
\boldsymbol{\sigma}(\theta, \phi):=(R \sin \theta \cos \phi, R \sin \theta \sin \phi, R \cos \theta) \;,
\end{eqnarray}
except for the north and south pole and the arc connecting them along $\theta=0$. The integral of a smooth function over the sphere $S^2$ can then be computed only by considering the single chart since this arc has no area. Here and subsequently, we sometimes write $\boldsymbol{\sigma}$ instead of $\boldsymbol{\sigma}(\theta, \phi)$ for simplicity of notation. It is easy to integrate  the first factor on the right hand side of  (\ref{scheqboundstateinmomentum}) and get
\begin{eqnarray}
\int_{S^2} e^{-i \mathbf{p}\cdot \boldsymbol{\sigma}} R^2 d \Omega = \int_{0}^{2\pi} \int_{0}^{\pi} e^{-i p R \cos(\theta)} R^2 \sin(\theta) d \theta d \phi =  \frac{4 \pi R}{p} \sin(p R) \;.
\end{eqnarray}
Substituting this into (\ref{scheqboundstateinmomentum}), we obtain the formal square integrable solution (thanks to the Parseval theorem \cite{appel2007mathematics}, which simply states that if a function is square integrable in momentum space, it is square integrable in position space, and they are equal)
\begin{eqnarray}
\widehat{\psi}(\mathbf{p}) = \frac{\lambda}{p^2 + \nu^2} \frac{\sin(p R)}{p R} \left(\int_{S^2} \psi(\boldsymbol{\sigma}) R^2 d \Omega \right) \;. \label{boudnstatewavefuncmomentum}
\end{eqnarray}
This still includes the integration of the unknown bound state wave function restricted to the sphere. Nevertheless, it can be found by imposing the following consistency condition, that is, 
\begin{eqnarray}
\int_{S^2} \psi(\boldsymbol{\sigma}) R^2 d \Omega = \left(\int_{S^2} \psi(\boldsymbol{\sigma}) R^2 d \Omega \right)  \int_{S^2} \left( \int_{\mathbb{R}^3} \frac{\lambda }{p^2 + \nu^2} \frac{\sin(p R)}{p R}  e^{i \mathbf{p} \cdot \boldsymbol{\sigma}(\theta', \phi')} \; \frac{d^3 p}{(2\pi)^3} \right) R^2 d\Omega' \;.
\end{eqnarray}
Integrating over the angular variables, the above consistency condition yields
\begin{eqnarray}
1= \frac{2 \lambda}{\pi} \int_{0}^{\infty} \frac{\sin^2(p R)}{p^2 + \nu^2} \; d p  \;. 
\end{eqnarray}
Using the even extension of this integral and the trigonometric identity $\sin^2 (p R)=(1-\cos(2 p R))/2$, a simple application of the residue theorem \cite{brown2009complex} gives 
\begin{eqnarray}
\left(1-e^{-2 R \nu} \right) = \frac{2 \nu}{\lambda}  \;.  \label{boundstateenergytrans}
\end{eqnarray}
This transcendental equation has always one solution as long as the 
slope of the left hand side is less than the slope of the right hand side around $\nu=0$. This gives us the necessary condition for the existence of a single bound state, that is, 
\begin{eqnarray}
\frac{1}{\lambda R} <1 \;. \label{boundstatecondition}
\end{eqnarray}
One can also write explicitly the bound state energy in terms of the Lambert $W$ function \cite{veberivc2012lambert}
\begin{eqnarray}
E= - \bigg(\frac{\lambda}{2} + \frac{1}{2R} W \left( - \lambda R e^{- \lambda R}\right)\bigg)^2 \;,
\end{eqnarray}
where $W$ is defined by the solution $x=W(c)$ of equation $ x e^{x}= c$. 

The  wave function corresponding to  this bound state is just the inverse Fourier transformation of (\ref{boudnstatewavefuncmomentum}) 
\begin{eqnarray}
\psi(\mathbf{r})= N \; \frac{\lambda}{2 \pi^2 r R} \int_{0}^{\infty} \frac{\sin(p R) \sin(p r)}{p^2 + \nu^2} d p \;,
\end{eqnarray}
where $\nu=\frac{\lambda}{2} + \frac{1}{2R} W \left( - \lambda R e^{- \lambda R}\right) $ and the surface integral of the wave function is absorbed into the normalization constant $N$. Since the integrand is an even function of $p$, and $\sin(p R) \sin(pr)= \frac{1}{2} \left( \cos(p(r-R))-\cos(p(r+R))\right)$ we can compute the integral by  means of  the residue theorem. We have two simple poles at $p=\pm i \nu$. For the case $r>R$ the contour consists of two parts, one along the real axis $(-\rho, \rho)$, and the other one is just the semi-circle of radius $\rho$ enclosed from the upper half-plane in the complex $p$ plane. For the case $0<r<R$, we must close the contour from the lower half-plane by a semi-circle. After taking the limit $\rho \rightarrow \infty$, we evaluate the above integral and obtain
\begin{equation}
\psi(\mathbf{r})= \begin{cases} 
      \frac{N \lambda}{4\pi r R \nu} e^{-\nu R} \sinh (\nu r) & 0< r \leq R \;, \\ \\
      \frac{N \lambda}{4\pi r R \nu} e^{-\nu r} \sinh (\nu R) & r \geq R  \;.
      \end{cases} \label{bswavefunction3d}
\end{equation}
It is worth pointing out that the derivative of the above bound state wave function with respect to $r$ has a discontinuity at $r=R$.

\subsection{Stationary Scattering Problem}

For the stationary scattering problem of the delta shell potential, we need to solve the following equation 
\begin{eqnarray} \label{algebraiceq3d}
(p^2-k^2) \widehat{\psi}(\mathbf{p}) = \alpha(p)\;,
\end{eqnarray}
where\footnote{Here again we need to make sure that the scattering solutions we are after are actually integrable over the sphere, the approach we take at this point is to assume so, and after finding the solution we check that this is a consistent assumption}. 
\begin{eqnarray}
\alpha(p):= \frac{\lambda}{p R} \sin(p R) \left(\int_{0}^{2\pi} \int_{0}^{\pi} \psi(\boldsymbol{\sigma}(\theta, \phi)) R^2 \sin \theta d \theta d \phi \right) \;.
\end{eqnarray}
The distributional solution of the algebraic equation (\ref{algebraiceq3d}) is obtained in the same manner as in the one dimensional case
\begin{eqnarray}
\widehat{\psi}(\mathbf{p}) = A \delta(\mathbf{p}-\mathbf{k}) + B \delta (\mathbf{p}+\mathbf{k}) + C \delta(p-k) + \mathrm{pv} \left(\frac{\alpha(p)}{p^2-k^2}\right) \;.
\end{eqnarray}
The only difference here is to take into account of all homogenous solutions since $(p^2-k^2) \delta(\mathbf{p} \pm \mathbf{k})=0$ as well as $(p^2-k^2) \delta(p - k)=0$. Due to the  positivity of the variables $p$ and $k$ we must disregard a possible  term of the form  $\delta(p+k)$. Then, the position space wave function can be obtained by taking the formal inverse Fourier transformation of the above distributional solution
\begin{eqnarray} \label{scatteringsolution3dstep1}
\psi(\mathbf{r})= \frac{A}{(2\pi)^3} e^{i \mathbf{k} \cdot \mathbf{r}} + \frac{B}{(2\pi)^3} e^{-i \mathbf{k} \cdot \mathbf{r}} + \frac{2 C k}{(2\pi)^2} \; \frac{\sin(k r)}{r} + \mathrm{pv} \int_{\mathbb{R}^3} \frac{e^{i \mathbf{p}\cdot \mathbf{r}} \alpha(p)}{p^2-k^2} \; \frac{d^3 p}{(2\pi)^3} \;.
\end{eqnarray}
The principal value of an integral over the space $\mathbb{R}^3$ can be computed by evaluating the principal value of the function inside the integral with respect to the radial coordinate $p$ (see e.g., \cite{appel2007mathematics}). Then, we get
\begin{eqnarray}
\mathrm{pv} \int_{\mathbb{R}^3} \frac{e^{i \mathbf{p}\cdot \mathbf{r}} \alpha(p)}{p^2-k^2} \; \frac{d^3 p}{(2\pi)^3} = \frac{2\lambda}{(2\pi)^2 r R} \; \left( \int_{S^2} \psi(\boldsymbol{\sigma}) R^2 d \Omega \right)\; \mathrm{pv} \int_{0}^{\infty} \frac{\sin(p r) \sin(p R)}{p^2-k^2} \; d p \;,
\end{eqnarray}
where we have used $d^3 p=p^2 d p \sin (\theta) d \theta d \phi$ and integrated over the angle variables. Using the trigonometric identity $\sin(p R) \sin(pr)= \frac{1}{2} \left( \cos(p(r-R))-\cos(p(r+R))\right)$, and write the each cosine term as the real part of complex exponential, we have   
\begin{align}
    \mathrm{pv} \int_{\mathbb{R}^3} \frac{e^{i \mathbf{p}\cdot \mathbf{r}} \alpha(p)}{p^2-k^2} \; \frac{d^3 p}{(2\pi)^3} = \frac{\lambda}{2 (2\pi)^2 r R} \; & \left( \int_{S^2} \psi(\boldsymbol{\sigma})  R^2 d \Omega \right)\; \nonumber \\ & \times \Bigg[ \mathrm{Re} \bigg(\mathrm{pv} \int_{-\infty}^{\infty} \frac{e^{ip(r-R)}}{p^2-k^2} d p \bigg) - \mathrm{Re}\bigg(\mathrm{pv} \int_{-\infty}^{\infty} \frac{e^{ip(r+R)}}{p^2-k^2} d p  \bigg) \Bigg]  \;.
\end{align}
Since 
\begin{align}
    \mathrm{pv} \int_{-\infty}^{\infty} \frac{e^{ip x}}{p^2-k^2} d p = \frac{1}{2k} \Bigg[  \mathrm{pv} \int_{-\infty}^{\infty} \frac{e^{ip x}}{p-k} d p - \mathrm{pv} \int_{-\infty}^{\infty} \frac{e^{ip x}}{p+k} d p \Bigg] \;,
\end{align}
and using the result for the principal values obtained in the one dimensional problem, we obtain in the scattering region ($r>R$) 
\begin{eqnarray}
\mathrm{pv} \int_{\mathbb{R}^3} \frac{e^{i \mathbf{p}\cdot \mathbf{r}} \alpha(p)}{p^2-k^2} \; \frac{d^3 p}{(2\pi)^3} = \frac{\lambda}{4\pi r R k} \sin(k R) \cos(k r) \left(\int_{S^2} \psi(\boldsymbol{\sigma}) R^2 d \Omega \right) \;.
\end{eqnarray}
Substituting back this into the formal scattering solution (\ref{scatteringsolution3dstep1}), we get
\begin{eqnarray}
\psi(\mathbf{r})= \frac{A}{(2\pi)^3} e^{i \mathbf{k} \cdot \mathbf{r}} + \frac{B}{(2\pi)^3} e^{-i \mathbf{k} \cdot \mathbf{r}} + \frac{2 C k}{(2\pi)^2} \; \frac{\sin(k r)}{r} + \frac{\lambda}{4\pi r R k} \sin(k R) \cos(k r) \left(\int_{S^2} \psi(\boldsymbol{\sigma}) R^2 d \Omega \right)  \;.
\end{eqnarray}
This is still a formal solution since the right hand side includes the unknown scattering wave function $\psi$. However, one can solve it by simply imposing the consistency condition, that is, we integrate both sides with respect to the measure on the sphere $R^2 d\Omega$ and deduce
\begin{eqnarray}
\int_{S^2} \psi(\boldsymbol{\sigma}) R^2 d \Omega = \left( \frac{A+B}{2\pi^2 k} + \frac{2 C k}{\pi}\right) \left(1-\frac{\lambda}{2k}\sin(2 k R) \right)^{-1} R \sin(k R) \;.
\end{eqnarray}
Putting this result into the formal scattering solution $\psi(\mathbf{r})$, we finally obtain
\begin{align}
\psi(\mathbf{r}) =  \frac{A}{(2\pi)^3} e^{i \mathbf{k} \cdot \mathbf{r}} + \frac{B}{(2\pi)^3} e^{-i \mathbf{k} \cdot \mathbf{r}}  + \frac{2 C k}{(2\pi)^2} \; \frac{\sin(k r)}{r} & + \frac{\lambda}{4\pi k} \sin^2(k R)  \nonumber \\ & \times \left( \frac{A+B}{2\pi^2 k} + \frac{2 C k}{\pi}\right) \left(1-\frac{\lambda}{2k}\sin(2 k R) \right)^{-1} \frac{\cos(k r)}{r} \;.
\end{align}
Let us now apply the outgoing boundary condition (\ref{outgoingboundaryconditions}) in three dimensions.
Therefore, we need to impose the following three simultaneous conditions 
\begin{eqnarray}
B & = & 0 \\
\frac{2 C k}{(2\pi)^2}  & = & i \beta \\ \frac{\lambda}{4\pi k} \sin^2(k R) \left( \frac{A+B}{2\pi^2 k} + \frac{2 C k}{\pi}\right) \left(1-\frac{\lambda}{2k}\sin(2 k R) \right)^{-1} & = & \beta \;,
\end{eqnarray}
for some arbitrary complex number $\beta$. One can express $\beta$ in terms of the coefficient $A$ by eliminating $C$ from the above equation and find
\begin{eqnarray}
\beta = \frac{A}{(2\pi)^3} \frac{\sin^2(k R)}{k^2} \left( \frac{1}{\lambda} + \frac{1}{2 i k} \left(1-e^{2i k R}\right)\right)^{-1} \;,
\end{eqnarray}
where we have used the trigonometric identity $\sin^2(kR)=\frac{1-\cos(2 k R)}{2}$. Hence, we obtain the scattering solution as
\begin{equation}
    \psi(\mathbf{r}) =  \frac{A}{(2\pi)^3} \Bigg[ e^{i \mathbf{k} \cdot \mathbf{r}} +  \frac{\sin^2(k R)}{k^2} \left( \frac{1}{\lambda} + \frac{1}{2 i k} \left(1-e^{2i k R}\right)\right)^{-1} \frac{e^{i k r}}{r}\Bigg] \;,
\end{equation}
which is exactly the same result obtained by considering the partial wave analysis and then restricting the problem to the $s$ wave sector ($l=0$) \cite{griffiths2016introduction}, see Appendix A for the details.

\section{Circular Dirac Delta Potential in Two Dimensional Momentum Space}
\label{Circular Dirac Delta Potential in Momentum Space}

\subsection{Bound State Problem}

We now consider the circular Dirac delta potential in two dimensions,
\begin{eqnarray}
H=H_0 - \lambda | \delta_{S^1} \rangle \langle \delta_{S^1}| \;, \label{Hamiltoniancircle}
\end{eqnarray}
where the circle $S^1$ centered at the origin with radius $R$ is parametrized by $\boldsymbol{\gamma}(\theta):= (R \cos (\theta), R \sin (\theta))$ and %
\begin{eqnarray}
\langle \delta_{S^1}| \psi \rangle = \frac{1}{\sqrt{L(S^1)}} \int_{0}^{2 \pi} \psi(\boldsymbol{\gamma}(\theta)) R d \theta \;,
\end{eqnarray}
with $L(S^1)$ is the length of the circle\footnote{For square integrable solutions, as we remarked in the sphere case, one can actually justify that the integral over the circle makes sense}. Using the generalized closure relation $\mathrm{Id} =\int_{\mathbb{R}^2} | \mathbf{r} \rangle \langle \mathbf{r}| d^2 r$, the above definition implies that
\begin{eqnarray}
\langle \mathbf{r} | \delta_{S^1} \rangle = \frac{1}{\sqrt{L(S^1)}} \int_{0}^{2 \pi} \delta(\mathbf{r}-\boldsymbol{\gamma}(\theta)) R d \theta \;. \label{rdeltacircle}
\end{eqnarray}
The time-independent Schr\"{o}dinger equation for the above Hamiltonian (\ref{Hamiltoniancircle}) in momentum space becomes
\begin{eqnarray}
\langle \mathbf{p} | H_0| \psi \rangle - \lambda \langle \mathbf{p} | \delta_{S^1} \rangle \langle \delta_{S^1}|\psi\rangle = -\nu^2  \widehat{\psi}(\mathbf{p}) \;. \label{scheqcircle}
\end{eqnarray}
Using (\ref{rdeltacircle}) in $\langle \mathbf{p}| \delta_{S^1} \rangle$, Equation (\ref{scheqcircle}) can be written as
\begin{eqnarray}
(p^2 + \nu^2) \widehat{\psi}(\mathbf{p}) = \frac{\lambda}{L(S^1)} \left( \int_{0}^{2\pi} e^{-i \mathbf{p} \cdot \boldsymbol{\gamma}(\theta)} R d \theta \right) \left( \int_{0}^{2\pi} \psi(\boldsymbol{\gamma}(\theta)) R d \theta \right) \;.
\end{eqnarray}
The first integral on the right hand side is 
\begin{eqnarray}
\int_{0}^{2\pi} e^{-i \mathbf{p} \cdot \boldsymbol{\gamma}(\theta)} R d \theta  = \int_{0}^{2\pi} e^{-i p R \cos(\theta)} R d \theta = 2 \pi R J_0(p R) \;,
\end{eqnarray}
thanks to the integral representation of the Bessel function $J_0(x)$, given by \cite{arfken2005mathematical}
\begin{eqnarray}
J_0(x)= \frac{1}{2 \pi} \int_{0}^{2\pi} e^{i x \cos(\theta)} d \theta \;.
\end{eqnarray}
Then, we obtain the formal solution of the Fourier transformed bound state wave function as
\begin{eqnarray}
\widehat{\psi}(\mathbf{p}) = \frac{\lambda J_0(p R)}{p^2 + \nu^2} \left( \int_{0}^{2\pi} \psi(\boldsymbol{\gamma}(\theta)) R d \theta \right) \;. \label{fouriertransformedBSwavefunctioncircle}
\end{eqnarray}
By following a similar approach as in the previous case, we evaluate the  integral of (the inverse Fourier transform of) the wave function (\ref{fouriertransformedBSwavefunctioncircle}) over the circle and get our consistency condition
\begin{eqnarray}
\frac{1}{\lambda R} = \int_{0}^{\infty} \frac{J_{0}^{2}(p R)}{p^2 + \nu^2} \; p dp \;, \label{boundstateequationcircle}
\end{eqnarray}
where we  found the integral over the angle variable $\theta$. Using the result (in page 671 in \cite{gradshteyn2014table})
\begin{eqnarray}
\int_{0}^{\infty} \frac{x J_{0}^{2}(x)}{x^2 + a^2} d x = I_{0}(a) K_0(a) \;,
\end{eqnarray}
where $I_0$ and $K_0$ are modified Bessel functions of the first and second kind, respectively, we obtain for the consistency equation
\begin{eqnarray}
\frac{1}{\lambda R}= I_0(\nu R) K_0(\nu R) \;. \label{finalbsenergycircle}
\end{eqnarray}
In contrast to the sphere case, we can not solve this transcendental equation analytically. However, there is a unique solution $\nu$ for given $\lambda$ and $R$. This can be seen easily by simply going back to the integral on the right hand side of Equation (\ref{boundstateequationcircle}) and taking the derivative of it with respect to $\nu$ under the integral sign and obtain
\begin{eqnarray}
\frac{d}{d \nu} \int_{0}^{\infty} \frac{J_{0}^{2}(p R)}{p^2 + \nu^2} \; p dp = - \int_{0}^{\infty} \frac{2 \nu J_{0}^{2}(p R)}{(p^2 + \nu^2)^2} \; p dp < 0 \;.
\end{eqnarray}
This shows that the right hand side of (\ref{finalbsenergycircle}) is a monotonically decreasing function whereas the left hand side is constant so that the solution always exists whatever the values of $\lambda$ and $R$ are.

Let us call the solution of Equation (\ref{finalbsenergycircle}) as $\nu_{*}$, then the bound state energy is symbolically expressed by $E=-\nu_{*}^2$. The bound state wave function associated with this bound state energy can then be found by taking the inverse Fourier transform of the solution (\ref{fouriertransformedBSwavefunctioncircle}). For this, we employ the following integral,  (see page 672 in \cite{gradshteyn2014table}) 
\begin{eqnarray}
\int_{0}^{\infty} \frac{J_0(a x) J_0(b x)}{x^2 + c^2} x dx = \begin{cases}
I_0(bc) K_0(ac)& \mathrm{if} \; 0<b<a  \\ \\
I_0(ac) K_0(bc)& \mathrm{if} \; 0< a< b \;,
\end{cases}\label{integbessel2}
\end{eqnarray}
as a result, we find the bound state wave function in terms of the solution $\nu_*$ up to a normalization constant $N$. 
\begin{eqnarray}
\psi(\mathbf{r}) = N \begin{cases}
I_0(\nu_{*} r) K_0(\nu_{*} R)& \mathrm{if} \; 0<r \leq R \\ \\
I_0(\nu_* R) K_0(\nu_* r)& \mathrm{if} \; r \geq R \;.
\end{cases}
\end{eqnarray}

\subsection{Stationary Scattering Problem}

A general distributional solution to the algebraic equation for the time-independent Schr\"{o}dinger equation in momentum space for the scattering problem, where $E=k^2$  is similarly given by\footnote{Here, the square integrability of the scattering wave function over the circle is an assumption to be verified after the solution is found}
\begin{eqnarray}
\widehat{\psi}(\mathbf{p}) = A \delta(\mathbf{p}-\mathbf{k}) + B \delta(\mathbf{p}+ \mathbf{k}) + C \delta(p-k) + \left( \int_{0}^{2\pi} \psi(\boldsymbol{\gamma}(\theta)) R d \theta \right) \mathrm{pv} \left( \frac{\lambda J_0(p R)}{p^2-k^2}\right) \;.
\end{eqnarray}
Taking the inverse Fourier transformation of this, we get
\begin{eqnarray}
\psi(\mathbf{r}) = \frac{A}{(2\pi)^2} e^{i \mathbf{k} \cdot \mathbf{r}} + \frac{B}{(2\pi)^2} e^{i \mathbf{k} \cdot \mathbf{r}}  + \frac{C k}{2\pi} J_0(kr) + \left( \int_{0}^{2\pi} \psi(\boldsymbol{\gamma}(\theta)) R d \theta \right) \mathrm{pv} \int_{\mathbb{R}^2} \left( \frac{\lambda J_0(p R)}{p^2-k^2}\right) e^{i \mathbf{p} \cdot \mathbf{r}} \frac{d^2 p}{(2\pi)^2} \;.
\end{eqnarray}
The principal value here can be computed by the analytic continuation of the result (\ref{integbessel2}) in the variable $c=-i k$ for the scattering region $r>R$ so that we find
\begin{eqnarray}
\psi(\mathbf{r}) = \frac{A}{(2\pi)^2} e^{i \mathbf{k} \cdot \mathbf{r}} + \frac{B}{(2\pi)^2} e^{i \mathbf{k} \cdot \mathbf{r}}  + \frac{C k}{2\pi} J_0(kr) + \left( \int_{0}^{2\pi} \psi(\boldsymbol{\gamma}(\theta)) R d \theta \right) \frac{i \lambda}{4} J_0(k R) H_{0}^{(1)}(kr) \;, \label{scatsolcircle1}
\end{eqnarray}
where $H_{0}^{(1)}$ is the Hankel function of the first kind  and $K_0(-ik r)= \frac{i \pi}{2} H_{0}^{(1)}(kr)$ \cite{lebedev1965special}.
In order to find the unknown factor in the bracket, we take the line integral of both sides over the circle to get
\begin{eqnarray}
\left( \int_{0}^{2\pi} \psi(\boldsymbol{\gamma}(\theta)) R d \theta \right) = J_0(k R) \left(\frac{(A+B)R}{2\pi}+ C k R\right) \left(1-\frac{i \pi \lambda R}{2} J_0(k R) H_{0}^{(1)}(k R)\right)^{-1} \;.
\end{eqnarray}
Substituting back this into (\ref{scatsolcircle1}) and using the asymptotic formulas for the Bessel functions \cite{lebedev1965special} 
\begin{eqnarray}
J_0(x) & \sim & \sqrt{\frac{2}{\pi x}} \cos(x-\frac{\pi}{4}) \\ 
H_{0}^{(1)}(x) & \sim & \sqrt{\frac{2}{\pi x}} e^{i(x-\frac{\pi}{4})} \;,
\end{eqnarray}
as $x \rightarrow \infty$, we obtain the asymptotic form of the scattering wave function 
\begin{align}
\psi(\mathbf{r}) & = \frac{A}{(2\pi)^2} e^{i \mathbf{k} \cdot \mathbf{r}} + \frac{B}{(2\pi)^2} e^{i \mathbf{k} \cdot \mathbf{r}}  + \frac{C k}{2\pi} \sqrt{\frac{2}{\pi k r}} \cos(k r-\frac{\pi}{4}) \nonumber \\ & +  \left(\frac{(A+B)R}{2\pi}+ C k R\right) \left(1-\frac{i \pi \lambda R}{2} J_0(k R) H_{0}^{(1)}(k R)\right)^{-1} \frac{i \lambda}{4} J_{0}^{2}(k R) \sqrt{\frac{2}{\pi k r}} e^{i(k r-\frac{\pi}{4})} \;.
\end{align}
Using the outgoing boundary condition in two dimensions (\ref{outgoingboundaryconditions2d}), this forces us to choose $B=C=0$ so that we obtain the scattering solution 
\begin{eqnarray}
\psi(\mathbf{r}) = \frac{A}{(2\pi)^2} \left( e^{i \mathbf{k} \cdot \mathbf{r}} + R \sqrt{\frac{\pi}{2k}} \left(\frac{1}{\lambda}-\frac{i \pi R}{2} J_0(k R) H_{0}^{(1)}(k R)\right)^{-1}  J_{0}^{2}(k R) \frac{e^{i k (r+\pi/4)}}{\sqrt{r}}  \right) \;.
\end{eqnarray}

\section{Comments on the Formal Operator Approach}

We would like to make a few comments about  the shell potentials in the usual differential equation approach. In order to make everything well-defined, it would be convenient first to regularize the formally expressed  Hamiltonian $H$ which has a singular interactions. If we think of a regularized Hamiltonian $H_\epsilon$ which keeps the  spherical symmetry, then $[H_\epsilon , {\mathbf L}]=0$ implies that the Hilbert space can be decomposed as a direct sum of  sub-Hilbert spaces labeled by the index $l$, that is, ${\mathcal H}=\bigoplus_{l=0} {\mathcal H}_l$ and we have a restriction of the Hamiltonian to each  subspace $\mathcal{H}_l$. Note that here we are not thinking of the regularized Hamiltonian as a rank one perturbation (as yet). The restricted Hamiltonian in spherical delta shell potential case will look like 
\begin{equation}
    H_\epsilon^{l}=-{1\over r^2 }{\partial \over \partial r}\Big(r^2{\partial \over \partial r}\Big)+{l(l+1)\over r^2}-\lambda \delta_\epsilon(r-R) \;.
\end{equation}
We can now interpret the last term as a rank one perturbation as we have done in the operator approach and remove the regularization by taking the limit as $\epsilon \rightarrow 0^+$. Note that this means we have a self-adjoint extension for each  Hamiltonian restricted to a particular subspace, and this leads to the results in \cite{demiralp2003properties}, Whereas in our approach the full Hamiltonian is a rank one perturbation of the free Hamiltonian by $|\delta_{S^2} \rangle \langle \delta_{S^2}|$. Indeed if we actually use the spherical symmetry to write the wave functions by means of the spherical basis elements and apply our rank one perturbation, we have
\begin{equation}
    \Psi(r,\Omega)=\sum_{l=0} \chi_l(r) Y_{lm}(\Omega) \mapsto \int_{S^2} \sum_{l=0}\chi_l(R) Y_{lm} (\Omega) d\Omega=4\pi \chi_0(R) \;,
    \end{equation}
as a result, the operators on  $l\neq 0$ sectors become  just the free Laplacian.
Hence, we see that the formal differential equation approach is a {\it different realization} of the same problem which exploits the symmetry completely.
Indeed if one wants to generalize the problem to delta functions supported on arbitrary smooth surfaces, which do not exhibit any symmetries in general, we believe that the rank one perturbation interpretation of the $2$ or $3$-dimensional Laplacian is the most natural one.

\section{Conclusion}

We have solved the bound state and stationary scattering problems of circular and spherical delta shell potentials for the $l=0$ sector using a direct formal operator approach, where we have expressed the interaction  as a rank one projection operator of the form $|\delta_{S} \rangle \langle \delta_{S}|$, where $S$ is either circle in two dimensions or sphere in three dimensions. In contrast to the standard approach, known as the partial wave analysis in the literature, we obtain the same results directly using the distributional solutions in the scattering problem and use the boundary conditions explicitly instead of using them implicitly in $i\epsilon$ prescription.

\section*{Appendix A: Partial Wave Analysis of Spherical Delta Shell Potential} \label{partialwaveanalysis3d}

We briefly summarize the results for the bound state and the scattering solution of the spherical delta shell potential (\ref{deltashellpotential3d}) via partial wave analysis in the low energy limit, where it is sufficient to consider $l=0$ terms. It is well-known that if potential is spherically symmetric as above, we start with the separable solution of the time-independent Schr\"{o}dinger equation in the following form  $\psi(r,\theta, \phi)= F(r) Y_{l}^{m}(\theta, \phi)$, where $Y_{l}^{m}$ are spherical Harmonics \cite{griffiths2016introduction} and $F(r)$ satisfies the radial Schr\"{o}dinger equation,
\begin{eqnarray}
-\frac{d^2 F(r)}{d r^2} -\frac{2}{r} \frac{d F(r)}{d r} + \left(V(r) + \frac{l (l+1)}{r^2} \right)F(r) = E F(r) \;. \label{equationforR}
\end{eqnarray}
Let us first consider the bound state problem for $l=0$. Then, the solution to equation (\ref{equationforR}) for $E=-\nu^2$ gives
\begin{eqnarray} F(r)=
\begin{cases}
A \frac{e^{\nu r}}{r} + B \frac{e^{- \nu r}}{r} & \mathrm{if} \; r < R 
\\\\ 
C \frac{e^{-\nu r}}{r} & \mathrm{if} \; r > R \;.
\end{cases} \label{partialwavebswavefunc}
\end{eqnarray}
The regularity of the solution $R$ around $r=0$ implies that $B=-A$. The continuity condition at $r=R$ and the jump discontinuity condition at $r=R$ yield
\begin{eqnarray}
A(e^{\nu R}-e^{-\nu R}) & = & C e^{-\nu R} \\ A \nu (e^{\nu R}+e^{-\nu R}) & = & C(\lambda -\nu) e^{-\nu R} \;.
\end{eqnarray}
This system of equations can be easily solved and we obtain the same transcendental equation (\ref{boundstateenergytrans}) for the bound state energy. Using the above solution (\ref{partialwavebswavefunc}), the bound state wave function $\psi(\mathbf{r})=F(r)=u(r)/r$ gives the same result (\ref{bswavefunction3d}) up to normalization constant.

As for the scattering part of the problem, the solution of the above differential equation (\ref{equationforR}) for the given potential (\ref{deltashellpotential3d}) is given by
\begin{eqnarray}
F(r) = \begin{cases}
A \frac{\sin(k r)}{r} & \mathrm{if} \; r < R 
\\\\ 
B \frac{\sin(k r + \delta)}{r} & \mathrm{if} \; r > R \;,
\end{cases}
\end{eqnarray}
where $\delta$ is the phase shift to be determined. Here we have used the fact that the radial solution $F$ must be finite at the origin $r=0$. The continuity of the solution $F$ implies that 
\begin{eqnarray}
F(r)= B \frac{\sin(k R + \delta)}{\sin(k R)} \; \frac{\sin (k r)}{r} \;.
\end{eqnarray}
Using the jump discontinuity of this function $\frac{d F}{d r}\big|_{R^+}-\frac{d F}{d r}\big|_{R^-}=-\lambda F(r)$ at $r=R$ due to the delta shell function, we obtain
\begin{eqnarray}
\cot(\delta) = - \left( \cot(k R)- \frac{k}{\lambda \sin^2(k R)}\right) \;.
\end{eqnarray}
From the well-known formula \cite{griffiths2016introduction} for the $s$-wave scattering amplitude $f$ given by $f(\theta,k)= \frac{e^{i \delta}}{k} \sin(\delta)$ and rearranging the mathematical expressions to get
\begin{eqnarray}
f(\theta, k) = \frac{\sin^2(kR)}{k^2} \; \left(\frac{1}{\lambda}+\frac{1}{2i k} (1-e^{2i k R}) \right)^{-1} \;.
\end{eqnarray}

\section*{Appendix B: Partial Wave Analysis of Circular Delta Shell Potential} \label{partialwaveanalysis2d}

The radial part of the Schr\"{o}dinger equation in two dimensions for a potential depending only on the radial coordinate $r$ is given by 
\begin{eqnarray}
-\frac{d^2 F(r)}{dr^2} -\frac{1}{r} \frac{d F(r)}{d r} + \left(V(r) + \frac{m^2}{r^2} \right) F(r) = E F(r) \;. \label{equationforR2d}
\end{eqnarray}
In a similar fashion described for the three dimensional problem, we first consider the bound state problem for $m=0$. Then, the general solution to equation (\ref{equationforR2d}) for $E=-\nu^2$ is given by the Bessel functions 
\begin{eqnarray} F(r)=
\begin{cases}
A J_0(i \nu r) + B Y_0(i \nu r) & \mathrm{if} \; r < R 
\\\\ 
C J_0(i \nu r) + D Y_0(i \nu r)   & \mathrm{if} \; r > R \;.
\end{cases} \label{partialwavebswavefunc2d}
\end{eqnarray}
The regularity of the solution $R$ around $r=0$ implies that $B=0$ due to the asymptotic behaviour $Y_0(z) \sim \frac{2}{\pi} \log(z/2)$ as $z \rightarrow 0$ \cite{lebedev1965special}. Using $J_0(i \nu r)=I_0(\nu r)$ and the identity \cite{abramowitz1988handbook}
\begin{eqnarray}
Y_0(i \nu r)=i I_0(\nu r) - \frac{2}{\pi} K_0(\nu r) \;,
\end{eqnarray}
the above solution becomes
\begin{eqnarray}
F(r)=
\begin{cases}
A I_0(\nu r) & \mathrm{if} \; r < R 
\\\\ 
(C+i D) I_0(\nu r) - \frac{2 D}{\pi} K_0(\nu r)   & \mathrm{if} \; r > R \;.
\end{cases} \label{partialwavebswavefunc2d2}
\end{eqnarray}
For bound states, the wave function should decay as $r \rightarrow \infty$. Since $I_0(x) \sim \frac{e^{x}}{\sqrt{2 \pi x}}$  as $x \rightarrow \infty$ \cite{lebedev1965special}, we must have $D=i C$. Then, imposing the continuity and jump discontinuity of the function $R$ at $r=R$, and the Wronskian $W(I_0(z), K_0(z))=-1/z$ \cite{lebedev1965special} we obtain the same transcendental equation for the bound state energy (\ref{finalbsenergycircle}) that we have obtained.

For the scattering problem, the radial part of the Schr\"{o}dinger equation (\ref{equationforR2d}) can be solved  in two regions $r<R$ and $r>R$, respectively and the general solution for $E=k^2$ and $m=0$  is given by 
\begin{eqnarray}
F(r) = \begin{cases}
A J_0(k r) + B Y_{0} (k r) & \mathrm{if} \; r < R 
\\\\ 
B H_{0}^{(1)}(k r) + C H_{0}^{(2)}(k r) & \mathrm{if} \; r > R \;,
\end{cases}
\end{eqnarray}
where $H_{0}^{(1)}(k r)= J_{0}(k r) + i Y_0(k r)$ and $H_{0}^{(2)}(k r)= J_{0}(k r) - i Y_0(k r)$. Since $Y_0(kr)$ blows up near origin, we choose $B=0$. Using the asymptotic behaviour \cite{lebedev1965special}
\begin{eqnarray}
H_{0}^{(1)}(kr) & \sim & \sqrt{\frac{2}{\pi k r}} e^{i(k r-\frac{\pi}{4})} \;, \\ H_{0}^{(2)}(kr) & \sim & \sqrt{\frac{2}{\pi k r}} e^{-i(k r-\frac{\pi}{4})} \;,
\end{eqnarray}
the above solution in the scattering region consists of outgoing and incoming spherical waves. The conservation of probability implies that the amplitude of these waves must be the same up to a phase shift. For this reason, we have $C=B e^{-2 i \delta}$ where $\delta$ is the phase shift to be determined. Hence, we obtain
\begin{eqnarray}
F (r) & = &  \begin{cases}
A J_0(k r) & \mathrm{if} \; r < R 
\\\\ 
2 D \left( \cos(\delta) J_0(k r) - \sin(\delta) Y_0(k r) \right)  & \mathrm{if} \; r > R \;,
\end{cases}
\end{eqnarray}
where $D=B e^{-i \delta} = C e^{i \delta}$. The continuity and the jump discontinuity of the solution $R(r)$ at $r=R$ give
\begin{eqnarray}
2 D \left(J_0(k R) \cos(\delta) - Y_0(k R) \sin(\delta) \right) & = & A J_0(k R) \\ 2 D k \left( J_1(kR) \cos(\delta) - Y_1(k R) \sin(\delta) \right) & = & A\left( \lambda J_0(k R) + k J_1(k R) \right)  \;.
\end{eqnarray}
Using the Wronskian $W(J_0(z),Y_0(z))=2/(\pi z)$ \cite{lebedev1965special} we find the solutions 
\begin{eqnarray}
\cos(\delta) & = & \frac{A}{4 D} \left(2 + \pi R \lambda J_0(k R) Y_0(k R) \right) \\ \sin(\delta) & = & \frac{A \pi R \lambda}{4 D} J_{0}^{2}(k R) \;.
\end{eqnarray}
Since $e^{2 i \delta} = \frac{1 + i \tan(\delta)}{1 - i \tan(\delta)}$, we find the scattering amplitude 
\begin{eqnarray}
f(\theta) = R \sqrt{\frac{\pi}{2 k}} \left( \frac{1}{\lambda} - \frac{i \pi R}{2}J_0(k R) H_{0}^{(1)}(k R)\right)^{-1} J_{0}^{2}(k R) e^{i \pi/4} \;. \label{scatteringamplitude2dfinalresult}
\end{eqnarray}
Here we have used the relation between the scattering amplitude and the phase shift in two dimensions \cite{lapidus1982quantum, adhikari1986quantum, landau2013quantum}  
\begin{eqnarray}
f(\theta)= (e^{2 i \delta}-1) \frac{e^{-i \pi/4}}{\sqrt{2 \pi k}} \;.
\end{eqnarray}
The result (\ref{scatteringamplitude2dfinalresult}) is consistent with our result obtained obtained before.

\section*{Appendix C: Distributional Solutions of Equation $(x^2-a^2)T(x)=1$}

Consider first the homogenous part of equation
\begin{eqnarray}
(x^2-a^2) T(x)=0 \;. \label{homogenousequation}
\end{eqnarray}
Then, thanks to the property of Dirac delta distributions $\delta(x-a)f(x)=f(a) \delta(x-a)$, it follows that $\delta(x-a)$ and $\delta(x+a)$ are solutions of  (\ref{homogenousequation}). One can show that there are no regular distributional solutions to (\ref{homogenousequation}) since $\langle (x \pm a) T(x), \psi \rangle = \int_{\mathbb{R}} (x\pm a) f(x) \psi(x) d x =0$ for all test functions $\psi$ implies that $f(x)=0$ identically. In other words, there are no non-trivial regular distributional solution satisfying (\ref{homogenousequation}). Actually, there are also no other singular distributions rather than Dirac delta functions (see Ref. \cite{kanwal1998generalized}). 

The second step is to show that the principal value distribution satisfies $(x-a) \mathrm{pv}(\frac{1}{x-a})=1$. This can be seen from the definition of the principal value, that is,
\begin{eqnarray}
\langle (x-a) \mathrm{pv}\left(\frac{1}{x-a}\right), \psi \rangle = \lim_{\epsilon \rightarrow 0^+} \int_{|x-a|\geq \epsilon} \psi(x)  dx = \langle 1, \psi \rangle \;,
\end{eqnarray}
for all test function $\psi$. We now turn to the inhomogenous case 
\begin{eqnarray}
(x-a)T(x)=1 \;. \label{inhomogenouseq}
\end{eqnarray}
It follows easily from the above results that the solution of the above equation is given by $T(x)= \mathrm{pv} \left( \frac{1}{x-a}\right) + A \delta(x-a)$, where $A$ is arbitrary complex number. Therefore, the general solution to $(x^2-a^2)T(x)=1$ is immediately  obtained thanks to $(x^2-a^2) \delta(x-a)=0$ and $(x^2-a^2) \delta(x+a)=0$, and the principal value of $\frac{1}{x^2-a^2}$ can be decomposed in terms of the linear combination of the principal values of $\frac{1}{x-a}$ and $\frac{1}{x+a}$.

\section*{Acknowledgments}

The authors gratefully acknowledge the many helpful discussions of O. Teoman Turgut during the preparation of the paper. We would also like to thank the anonymous reviewer whose comments improved this manuscript.

\end{document}